\documentclass[osajnl,twocolumn,showpacs,superscriptaddress,10pt]{revtex4-1}
\usepackage{amsmath,amssymb,graphicx,epstopdf}

\begin{document}

\title{Microscopic Lensing by a Dense, Cold Atomic Sample}

\author{Stetson Roof}
\author{Kasie Kemp}
\author{Mark Havey}\email[Corresponding Author:]{mhavey@odu.edu}
\affiliation{Old Dominion University, Department of Physics, Norfolk, Virginia 23529}

\author{I.M. Sokolov}
\affiliation{Department of Theoretical Physics, State Polytechnic
University, 195251, St.-Petersburg, Russia}
\affiliation{Institute for Analytical Instrumentation, Russian Academy of Sciences, 198103, St.-Petersburg, Russia}
\author{D.V. Kupriyanov}
\affiliation{Department of Theoretical Physics, State Polytechnic
University, 195251, St.-Petersburg, Russia}

\begin{abstract}
We demonstrate that a cold, dense sample of $^{87}$Rb atoms can exhibit a micron-scale lensing effect, much like that associated with a macroscopically-sized lens. The experiment is carried out in the fashion of traditional z-scan measurements but in much weaker fields and where close attention is paid to the detuning dependence of the transmitted light. The results are interpreted using numerical simulations and by modeling the sample as a thin lens with a spherical focal length.
\end{abstract}


\maketitle

The z-scan technique, pioneered by Sheik-Bahae $\emph{et al.}$ \cite{Sheik-Bahae}, has become, over the past 25 years, a staple for measuring optical properties of non-linear optical media. The premise of the approach involves tightly focusing a near-resonant laser beam onto a sample and scanning the sample's location across the focus. The transmitted light is then sent through an aperture and collected on a detector, where the amount of transmitted light describes the optical properties.  This method has been extended to cold-atomic samples \cite{Labeyrie1,Labeyrie2,Wang1,Wang2}  where the measurement process is quite similar, differing mainly by the fact that the beam focus is moved and the sample remains stationary. These works involved using large, dilute ($\sim$10$^{10}$atoms/cm$^3$) samples formed in magneto-optical traps and very high probe laser intensities to obtain information on the non-linear index-of-refraction. In other research \cite{Kemp,Sokolov,Pellegrino,Javanainen}, particular attention has been paid to a contrasting regime that deals with light transmission through small, micron-sized samples with high atomic density (10$^{13}$-10$^{14}$atoms/cm$^3$) and where the probing beam has an intensity much lower than the saturation intensity of the atomic transition of interest. This density and probe intensity regime is particularly interesting because of cooperative and collective optical effects that are predicted to exist at these densities \cite{Friedberg,Dicke,Anderson}. However, this research typically has been concerned with the absorptive properties of the sample and not the refractive ones that naturally accompany them.

In this letter we describe z-scan measurements on a small, dense, cold atomic sample in the regime of low light intensity. Here, absorption and refraction each play a significant role in the amount of light transmitted through the sample and collected by a detector. By comparing the experimental results with simulations and modeling the sample as an effective lens, we show that the atomic ensemble serves as a microscopic lens capable of focusing an incident laser beam on a readily controllable micron length scale. Knowledge of this effect is useful for understanding forward transmission and fluorescence measurements of dense atomic samples, where refraction may modify observed absorption properties.

\begin{figure}
\includegraphics[scale=.3]{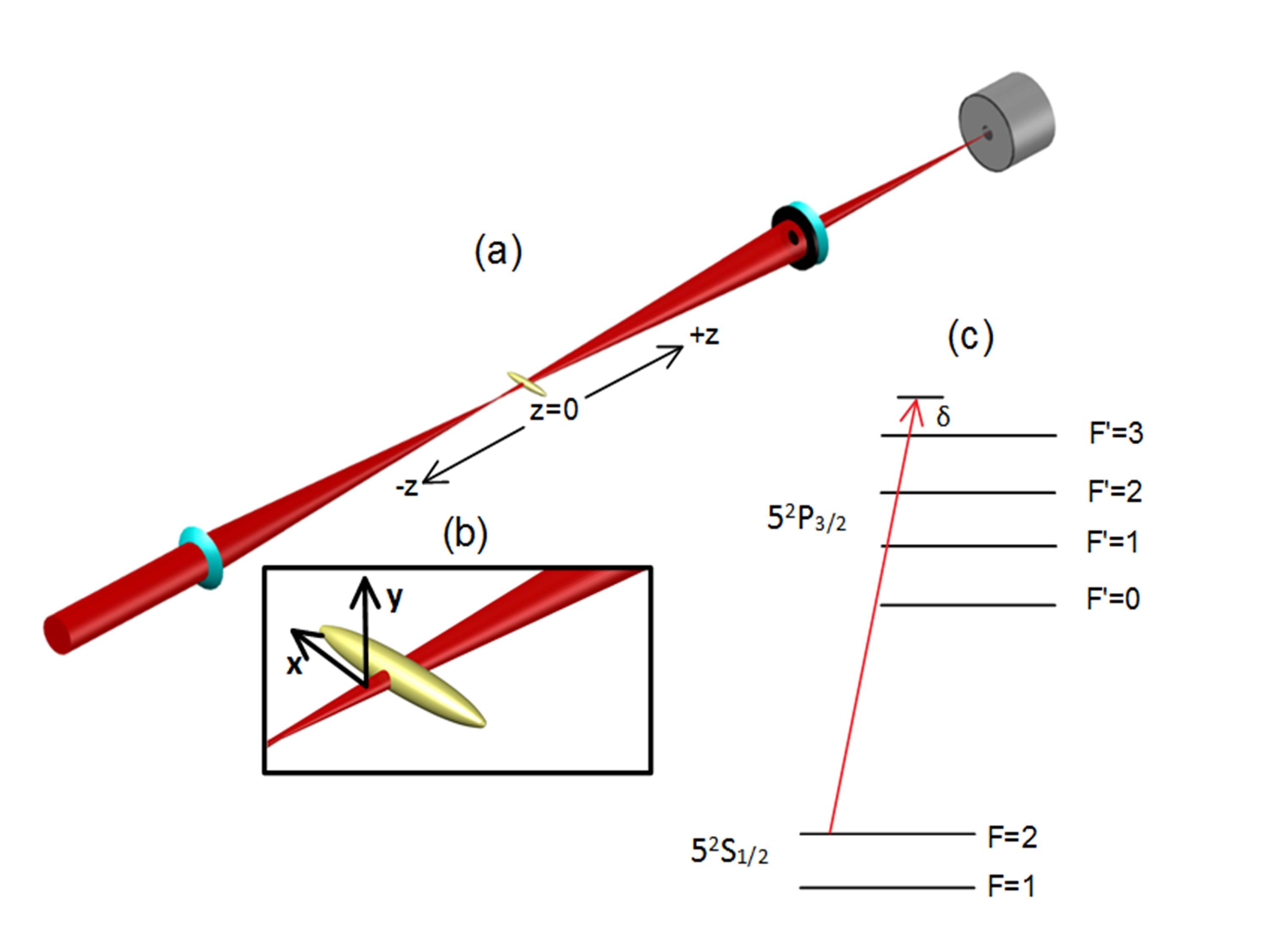}
\caption{(a) The experimental configuration used. A near-resonant light beam is incident on $^{87}$Rb atoms prepared in an optical dipole trap. The transmitted beam is apertured and collected on a CCD camera. Contrary to traditional z-scan measurements, the focal spot of the beam is moved rather than the sample. (b) An enhanced view of the incident probe on the sample, showing the coordinate system used. (c) Relevant hyperfine energy levels, showing the probe frequency scanned about the F=2 $\rightarrow$ F'=3 transition with a detuning $\delta$.}
\label{fig1}
\end{figure}

\begin{figure}
\includegraphics[scale=.5]{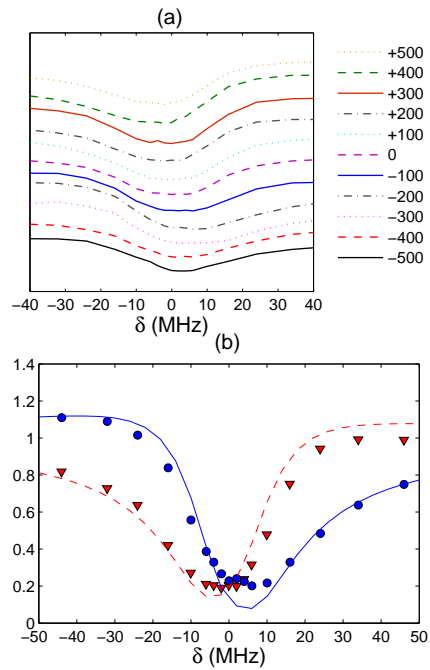}
\caption{(a) The frequency response for several different focal positions with respect to the atomic sample. Each curve is a step of 100 $\mu$m extending from z=-500 $\mu$m to z=+500 $\mu$m. To give a sense of contrast, each spectrum has been given a fixed offset from the previous one. (b) The normalized transmittance data for a focus at z=+200 $\mu$m (triangles) and z=-200 $\mu$m (circles) overlapped with the slip-step simulation.}
\label{fig2}
\end{figure}

The experimental setup is shown in Fig.~\ref{fig1}(a). There, a linearly polarized probe beam is focused to a 5 $\mu$m, 1/e$^2$ radius with an intensity of 300 $\mu$W/cm$^2$.  The beam bisects a sample of $^{87}$Rb atoms prepared in a far-off resonance optical dipole trap (FORT). The FORT, loaded from a standard magneto-optical trap (MOT), has a Gaussian profile given by
\begin{equation}
\rho(x,y,z)=\rho_0 \text{exp} \Big\{ -\frac{x^2}{2r_{x}^2}-\frac{(y^2+z^2)}{2r_{y}^2} \Big\}
\label{1}
\end{equation}
where $r_x$ and $r_y$ are $255$ $\mu$m and $3.5$ $\mu$m, respectively. Atoms in the trap, initially prepared in the 5$^2$S$_{1/2}$ F=1 ground state, have a temperature $\sim100$ $\mu$K and number around 9$\times$10$^5$, giving a peak density of 1.8$\times$10$^{13}$ atoms/cm$^{3}$. The experimental duty cycle consists of holding the atoms until they thermalize ($\sim$200 ms), upon which they are pumped to the higher energy F=2 ground level and probed about the resonant 5$^2$S$_{1/2}$ F=2 $\to$ 5$^2$P$_{3/2}$ F$'$=3 D2-line transition (see Fig.~\ref{fig1}(b)). The probe light is focused to controllable locations before and after the FORT by translating the focusing lens, and pulsed for a duration of 10 $\mu$s. This interval is chosen to maximize the signal-to-noise without introducing any significant optical pumping.  The transmitted light is then spatially filtered with a 0.5 cm diameter aperture and the resulting spatial distribution of intensity captured on a charge-coupled device camera (CCD). The aperture serves as a way to measure intensity and phase fluctuations by a saturable Kerr media, as has been done in other z-scan technique experiments \cite{Sheik-Bahae,Wang1}. For each detuning the probe signal is accumulated over 50 runs to acquire good signal-to-noise.
Shown in Fig.~\ref{fig2}(a) is a set of transmission spectra for each focal position of the probe beam.  The curves have been normalized by the unattenuated probe beam intensity ($\emph{i.e.}$ the probe transmission with no atomic sample present) and, for sake of demonstration, have been given a constant vertical-offset to show the signal progression as a function of z. It can be seen that there is a distinct $\emph{flip}$ in the frequency response of the sample depending on where the beam is focused. This can be attributed to focusing and defocusing effects similar to those described by Labeyrie $\emph{et al.}$ \cite{Labeyrie2}, except that in the present case the saturation parameter is very low (s$_0$ = 0.2). With observations as in Fig.~\ref{fig2}, it easy to perceive why absorption measurements of small, high density samples can be particularly cumbersome, especially when the numerical aperture of the imaging apparatus is small itself.

Measurements at high-field intensities (s$_0 \sim$500) were also performed; these yielded curves similar to the ones shown in Fig.~\ref{fig1}(a). The detailed shape of these curves, however, were dependent on the intensity.  We attribute this to the fact that, as the $F=2 \to F'=3$ transition is strongly saturated, off resonance scattering on the nearby $F=2 \to F'=1,2$ transitions contribute significantly to the spectral response. Modeling these effects is complicated, and would require calculation not only of the relative strength of these transitions, but  knowledge of the scaled cross-sections and line widths, which are generally known to be density dependent \cite{Kemp,Pellegrino}. These have not all been determined by measurement or calculation.  With this concern, and to maintain the cleanest interpretation of our measurements, we limited our focus to the low-field regime.

As a way to model the z-scan measurements, simulations were performed that closely resembled the experimental conditions. These simulations were based on the paraxial-wave-equation (PWE) for nonlinear media and implemented by the so-called `split-step' method; here a brief description is given. Under the conditions of the slowly-varying-envelope-approximation (SVEA), the complex amplitude of the electric field can be described by \cite{Boyd}
\begin{equation}
\frac{\partial \mathbf{A}}{\partial z}=\frac{i}{2k} \nabla_{\perp}^2 \mathbf{A}+\frac{ik}{2n_0^2} \chi \mathbf{A}
\label{2}
\end{equation}
where $\mathbf{A}$ is the complex electric field amplitude, $k=2\pi /\lambda$, $n_0$ is the linear index of refraction, and $\chi$ is the susceptibility given by (SI units)
\begin{equation}
\chi=-\frac{\sigma_0\rho(x,y,z)}{k_0} \frac{2\frac{\delta}{\gamma}-i}{1+4(\frac{\delta}{\gamma})^2+I/I_s}.
\label{3}
\end{equation}
Here, $\sigma_0$ is the resonant scattering cross-section, $\rho(x,y,z)$ is the spatially-dependent density, $k_0=2\pi/\lambda_0$ with $\lambda_0$ being the resonant wavelength of the transition, $I$ is the spatially-dependent intensity, and $I_s$ is the saturation intensity for the specific transition. Generally, the PWE must be solved numerically. One method, which is particularly simple and requires very little computational time, is the `split-step' algorithm. The split-step method, described more in detail in \cite{Newell}, involves propagating the beam forward in discrete steps in free-space and making non-linear corrections between each step. While the method is limited to computing the forward propagation of the beam, it serves well as a guide to understanding the experimental results. It should be noted that recent theoretical \cite{Sokolov,Bienaime} and experimental work \cite{Kemp} have shown that for a sufficiently high density the resonant cross-section should decrease with increasing density; this is accompanied by an increase in the transition linewidth. Therefore, $\sigma_0$ and $\gamma$ were varied in ($\ref{2}$) to match those obtained in \cite{Kemp} where a similar density was probed. Typical results for the simulation are shown in Fig.~\ref{fig2}(b) along with the corresponding experimental data for z=200 $\mu$m and -200 $\mu$m. For each simulated spectrum the light was collected at an effective distance of 10 times the Rayleigh length of the probe beam ($\sim$1 mm) away from the sample and on axis. The value of the collection distance was chosen so as to be in the 'far-field' regime and was limited in size by computational restraints. The comparison is good considering the scale of the atomic sample involved. The relatively small differences between the measurements and simulations could be due by uncertainty in the experimental-probe focus location as well as the propagation distance limitation in the simulation.

Physically speaking, our results can be described as a `lensing' effect \cite{Labeyrie1}, but on a much smaller length-scale than has been previously studied. This scale can be estimated by considering beam deflection along the smaller radius of the sample and approximating it as a ball lens with a focal length given by \cite{EdmundOptics}
\begin{equation}
f=\frac{n_0r_0}{2(n_0-1)}.
\label{4}
\end{equation}
Here $n_0$ is the index-of-refraction and $r_0$ is the radius of the lens, taken to be 2$r_y$, so that it is in terms of the beam profile definition. From the two-level atom susceptibility, the index-of-refraction can be written as
\begin{equation}
n_0 \approx 1+\frac{1}{2}\mathbf{Re}\{ \chi^{(1)} \}
\label{5}
\end{equation}
where $\chi^{(1)}$ is the linear electric susceptibility obtained by expanding Eq. (\ref{2}). Insertion of $\chi^{(1)}$ into (\ref{4}) gives
\begin{equation}
n_0=1-\frac{\sigma_0 \rho_0}{k_0} \frac{\frac{\delta}{\gamma}}{1+4(\frac{\delta}{\gamma})^2}
\label{6}
\end{equation}
Here the spatial dependence of the sample density is ignored. Observing the detuning dependence one can see that for $\delta < 0$, $n_0$ is greater than 1 leading to a positive focal length and hence a $\emph{focusing}$ effect. Likewise for $\delta > 0$, $n_0$ is less than 1 giving a negative focal length or $\emph{de-focusing}$ effect. To give an idea of scale for this ball-type  lens, the most acute response occurs for $\delta$=-$\gamma$/2 ($n_0$=1.02) and $\delta$=+$\gamma$/2 ($n_0$=0.98) which corresponds to $f$=+176 $\mu$m and $f$=-169 $\mu$m, respectively. With an effective focal length for the sample, a focusing distance can be estimated by using ray transfer matrices or the `ABCD' Law for Gaussian beams \cite{Verdeyen} which states that
\begin{equation}
\frac{1}{q_2}=\frac{C+D(1/q_1)}{A+B(1/q_1)}
\label{7}
\end{equation}
where A,B,C, and D are the elements for the transfer matrix and $q(z)$ is  the complex beam parameter given by
\begin{equation}
\frac{1}{q(z)}=\frac{1}{R(z)}-i\frac{\lambda}{\pi w^2(z)}
\label{8}
\end{equation}
Here $R(z)$ is the radius of curvature of the incident Gaussian beam and $w(z)$ is the beam waist. The transfer matrix can be constructed as a combination of a thin lens with focal length given by (\ref{3}) followed by a free-space propagation of length $z$
\begin{equation}
\mathbf{T}=
\begin{pmatrix}
1-\frac{z}{f} & z\\
\frac{-1}{f} & 1
\end{pmatrix}
\label{9}
\end{equation}
Solving for the minimum waist results in a focusing location of
\begin{equation}
z'=\frac{(1-\frac{f}{R(-z_0)})f}{(1-\frac{f}{R(-z_0)})^2+(\frac{f}{z_R'})^2},
\label{10}
\end{equation}
\noindent where $z_R'=\pi w^2(-z_0)/\lambda$ and $z_0$ is the intended probe focal location. This approach works well when the incident beam is comparable to or less than the size of the sample. Once the beam exceeds the dimensions of the sample, geometrical optics no longer applies; then diffraction makes a significant contribution to the propagated beam.
\begin{figure}
\includegraphics[scale=.5]{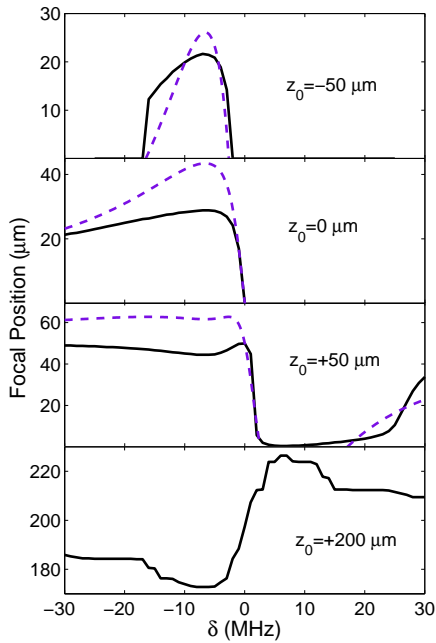}
\caption{Comparison between the model (dashed-line) and simulation (solid-line). $z_0$ is the \textit{intended} focus of the probe beam, but with the presence of the atom sample, the actual focal location is shifted along z. Due to limitations of the applicability of the model, only the simulation is shown in the lowest graph.}
\label{fig3}
\end{figure}

\begin{figure}
\includegraphics[scale=.45]{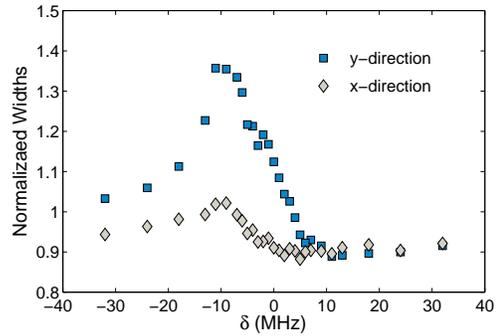}
\caption{The probe beam spatial intensity profile on the CCD camera with no aperture present and an intended probe focus of $z_0$ $>$ 0. With reference to  Fig.~\ref{fig1}(a), the spectra are labeled as the y and x direction referring to the FORT short and long axis directions, respectively. For each detuning, Gaussian radii are extracted from the CCD image and normalized to the beam radius with no sample present. For the y-direction of the atom sample there is more dispersion owing to a smaller radius of curvature. Even though the long-axis direction of the sample is much larger than the short-axis one, a slight distortion can still be seen showing the probe profile along that direction is also focused and defocused.}
\label{fig4}
\end{figure}

Using the above model and simulation, it becomes clear why the data takes the form it does in Fig.~\ref{fig2}(a). For an intended probe focus before the sample $z_0$, only the low frequency side of resonance affects the actual probe focus $z'$ by converging the beam after it passes through the sample (Fig.~\ref{fig3}). The high frequency side of resonance, due to having an index-of-refraction less than one, only serves to spread the beam out. This gives an absorption profile lower on the blue side than the red. However, once $z_0$ is sufficiently greater than zero, the red side of resonance will cause the probe beam to focus prematurely while the blue-side extends the focal location giving the illusion of more absorption for -$\delta$ (see the lowest curve in Fig.~\ref{fig3}). We may then consider the atomic sample as a microscopic lens as the probe beam can be caused to focus on a micron-length scale. This micro-lensing effect can be manifested in the far-field by observing the beam image with no aperture present, as in Fig.~\ref{fig4}. Here the probe beam passes through the sample and all of the light is collected on the CCD. From the beam image the relative amount of focusing or defocusing for each detuning can be seen by fitting the spatial profile as a bi-Gaussian and normalizing to the unattenuated probe beam profile. The changing spatial profile of the probe is due to the sample moving the effective focal location as a function of detuning.

In conclusion, measurements are presented that show that a micron-sized atomic sample can serve as an effective microscopic lens. These measurements are considerably different from previous z-scan experiments due to the relative size of the sample used and to the detailed concern with the frequency response of the sample. The data is interpreted by a beam propagation simulation that takes into account the non-linear response of the sample and further accounts for density dependent effects as determined in earlier experiments on similar samples \cite{Kemp}. Finally, a toy model is introduced that describes the macroscopic optical properties of the sample moderately well as a frequency dependent ball lens.

We appreciate financial support by the National Science Foundation (Grant No. NSF-PHY-1068159) and the Russian Foundation for Basic Research (Grant No. RFBR-CNRS 12-02-91056). We also thank Robert Boyd for an enlightening discussion about the z-scan method.

\end{document}